\documentstyle [11pt]{article}   %{report}
\def\ZzZ{{\hbox{\tenrm Z\kern-.31em{Z}}}}
\def\CcC{{\hbox{\tenrm C\kern-.45em{\vrule height.67em width0.08em depth-
.04em
\hskip.45em }}}}

\newcommand{\lab}{\label}

\newcommand{\bc}{\begin{center}}
\newcommand{\ec}{\end{center}}
\newcommand{\be}{\begin{equation}}
\newcommand{\ee}{\end{equation}}
\newcommand{\bea}{\begin{eqnarray}}
\newcommand{\eea}{\end{eqnarray}}
\newcommand{\bs}{\begin{subequations}}
\newcommand{\es}{\end{subequations}}
\newcommand{\beq}{\begin{eqalignno}}
\newcommand{\eeq}{\end{eqalignno}}

\newcommand{\half}{\frac{1}{2}}

\def\lab{\label}

\def\lab{\label}

\def\IiI{{\hbox{\tenrm I\kern-.19em{I}}}}

\def\uq2{U_q({\uit su}(2))}

\def\ggg{${\cal G} $}

\def\half{{1\over{2}}}

\def\CcC{{\hbox{\tenrm C\kern-.45em{\vrule height.67em width0.08em depth-.04em
\hskip.45em }}}}
\def\RrR{{\hbox{\tenrm I\kern-.17em{R}}}}
\def\HhH{{\hbox{\tenrm {I\kern-.18em{H}}\kern-.18em{I}}}}
\def\DdD{{\hbox{\tenrm {I\kern-.18em{D}}\kern-.36em {\vrule height.62em
width0.08em depth-.04em\hskip.36em}}}}

\def\ZzZ{{\hbox{\tenrm Z\kern-.31em{Z}}}}

\def\IiI{{\hbox{\tenrm I\kern-.19em{I}}}}
\def\NnN{{\hbox{\tenrm {I\kern-.18em{N}}\kern-.18em{I}}}}
\def\QqQ{{\hbox{\tenrm {{Q\kern-.54em{\vrule height.61em width0.05em
depth-.04em}\hskip.54em}\kern-.34em{\vrule height.59em width0.05em depth-.04em}}
\hskip.34em}}}
\def\OoO{{\hbox{\tenrm {{O\kern-.54em{\vrule height.61em width0.05em
depth-.04em}\hskip.54em}\kern-.34em{\vrule height.59em width0.05em depth-.04em}}
\hskip.34em}}}

\def\uq2{U_q({\uit su}(2))}

\def\fraz#1#2{{\strut\displaystyle #1\over\displaystyle #2}}

\def\part#1{\fraz{\partial}{\partial#1}}

\def\su2q{SU(2)_q}
\def\h1q{H(1)_q}

\def\nu{N_{1}}

\begin{document}

$$ $$

\bc
{
%\hspace{1cm}
{\bf HOW TO DEAL WITH THE ARROW OF TIME\\
IN QUANTUM FIELD THEORY}

$$ $$
%$$ $$

%\hspace{1cm}

 GIUSEPPE VITIELLO

\bigskip

%\hspace{1cm}

Dipartimento di Fisica, Universit\`a di Salerno, 84100
Salerno, Italy\\
INFM, Sezione di Salerno\\
INFN, Gruppo Collegato di Salerno\\
e-mail: vitiello@sa.infn.it

$$ $$
}

\ec

%\hspace{1cm}
{\bf Abstract}
The formalism of Quantum Mechanics
is based by definition on conserving
probabilities and thus there is no room for the description of
dissipative systems
in Quantum Mechanics. The treatment of time-irreversible evolution
(the arrow of time) is therefore ruled out by definition in
Quantum Mechanics. In Quantum Field Theory it is, however, possible to
describe time-irreversible evolution by resorting to the existence of
infinitely many unitarily inequivalent representations of the
canonical commutation relations (ccr). In this paper I review such a
result by discussing the canonical
quantization of
the damped harmonic oscillator (dho), a prototype of dissipative systems.
The irreversibility of time evolution is expressed as tunneling among
the unitarily inequivalent representations.
Canonical quantization is shown to lead to time dependent SU(1,1)
coherent states. The exact action for the dho is derived in
the path integral formalism of the quantum Brownian motion
developed by Schwinger and by Feynman and Vernon. The doubling of
the phase-space degrees of freedom for dissipative systems is
related to quantum noise effects.
Finally, the r\^ole of dissipation in the quantum
model of the brain and the occurrence that the cosmological
arrow of time, the thermodynamical one and the biological one point into
the same direction are shortly mentioned.

$$  $$

$$   $$

{\bf 1. Introduction}

\bigskip

The formalism of
Quantum Mechanics (QM) is
based on conserving
probabilities. In principle, therefore, there is no room for the
description of time-irreversible evolution
({\it the arrow of time}) in QM.
One has to introduce some sort of generalized
quantum formalism in order to describe dissipative systems.
The
developments of the theory of unstable states going beyond
the Breit-Wigner treatment and other phenomenological approaches
have been frequently reported in the literature. See for example
refs. \cite{Chiu:1992gt}-\cite{Bohm:1980cc}.

Dissipative systems have been
analyzed in the path integral formalism by Schwinger \cite{8.}
and by Feynman and Vernon \cite{9.}  from the point of view of the
quantum theory for Brownian motion
and are of course a major topic in non-equilibrium
statistical mechanics and non-equilibrium Quantum Field Theory (QFT)
at finite
temperature \cite{10.}-\cite{17.}.

In this paper I report on the
results \cite{Celeghini:1989qc}-\cite{Iorio:1993jn} on dissipative
systems in quantum theory which show that QFT does allow a correct
treatment of the arrow of time provided the full set of unitarily
inequivalent (ui) representations of the canonical commutation
relations (ccr) is used. I show \cite{Celeghini:1998sy} that
the proper algebraic structure of QFT is the deformed
Hopf algebra \cite{25.,Celeghini:1991km}
and that
the doubling of
the phase-space degrees of freedom implied by such a structure is
related to quantum noise effects in the case of dissipative systems
\cite{Srivastava:1995yf}.

The microscopic theory
for a dissipative system
must include the details of the processes responsible for dissipation,
including quantum
effects. One may start since the beginning with a Hamiltonian that
describes the system, the bath and the system-bath interaction.
Subsequently, the description of the original dissipative system is
recovered by the reduced density matrix obtained by eliminating
the bath variables which originate the damping and the fluctuations.
The problem with dissipative systems in QM is indeed that
ccr are not preserved by time evolution
due to damping terms. The r\^ole of fluctuating forces is in fact the one
of preserving the canonical structure.

It is known since long time \cite{12.} that, at a classical level,
the attempt to derive from a variational
principle the equations of motion defining the dissipative system
requires the introduction of additional complementary
equations.

This latter approach has been pursued since several years also in
context of quantum theory. In refs. \cite{13.} and
\cite{Celeghini:1989qc}-\cite{Iorio:1993jn}
the quantization
of the damped harmonic oscillator (dho) has been studied by doubling the
phase-space degrees of freedom (see also \cite{DeFilippo:1977bk} for the study of
unstable particle in QFT).
The doubled degrees of freedom play the r\^ole of
the bath degrees of freedom. Let me observe that the canonical formalism is devised solely for closed systems, therefore in order to produce the canonical quantization of the damped oscillator, it is
necessary and sufficient to close the system, namely to "balance" the energy flux, the momentum exchange, etc.. For that task, and only for that task, we do not really need to know the details of the environment and not even the details of the system-environment coupling: therefore, for such a limited task, we may "simulate" the environment as a collection of oscillators whose k-modes match the k-modes of our damped oscillator. With such a choice
the environment, depicted as the system time-reversed double, is treated as the "effective" environment. Of course, in those cases in which such a crude simplification is not enough (we might be really interested in the details of the system-environment interface, for example) much more care is needed and the doubling picture is not enough.
In Sec. 2 I present the approach based on the system doubling.

I would like to stress that the analysis for dissipative systems and
the arrow of time presented in this paper should not be considered
to be something just formal. It is a real problem
the one of the description of open systems in a mathematically consistent
formalism in QFT. QFT is in fact the only available theoretical scheme
to describe high energy physics, as well as condensed matter physics,
and quantum systems are always open systems interacting with their
environment. It is true that in many cases the approximation of
treating them as closed systems is very useful and successful
for phenomenological computations, nevertheless there are many
cases in which dissipative effects and breakdown of time reversal
symmetry cannot be neglected. In these latter circumstances we
do need a reliable, mathematically consistent QFT formalism.

The approach here presented has revealed to be useful in several
applications of physical interest, ranging from unstable particles
\cite{DeFilippo:1977bk}, to coherence in quantum Brownian motion
\cite{Blasone:1998xt},
squeezed states in quantum optics
\cite{Celeghini:1989qc,Celeghini:1993jh,Celeghini:1991jw}, topologically
massive
theories in the infrared region in 2+1 dimensions \cite{Blasone:1996yh},
the Chern-Simons-like dynamics of Bloch electrons in solids
\cite{Blasone:1996yh},
and has features also common to two-dimensional gravity models
\cite{Cangemi:1996yz},
to the study of quantization arising from the loss of information
\cite{'tHooft:1999gk,Blasone:2001ew}, to the quantization of matter
in curved background \cite{Martellini:1978sm}.
Moreover, it has been applied \cite{Vitiello:1995wv,Alfinito:2000ck,37.}
to the study of the
memory capacity problem in the quantum model for the brain \cite{31.}.

It has been known \cite{13.} that in QM
time evolution of the dho
leads out of the Hilbert space of states; in other words, the QM
treatment of dho does not provide a unitary
irreducible representation of SU(1,1) \cite{14.}. To
cure these pathologies one must move to QFT,
where infinitely many unitarily inequivalent
representations of the ccr are allowed (in the infinite volume  or
thermodynamic limit). The reason for this is that
the set of the states of the damped oscillator splits
into  ui
representations (i.e. into disjoint {\it folia},
in the C*-algebra formalism)
each one representing the states of the system at time {\it t}:
the time irreversible evolution is described as
{\it tunneling} between  ui
representations. A remarkable feature of this description thus emerges:
{\it
at
microscopic level the irreversibility of time evolution (the
arrow of time) is expressed by the non unitary evolution across the
ui representations of the ccr.}

I remark that the nature of the ground states of the ui
representations is the one of the  SU(1,1) generalized
coherent states. Furthermore,
the squeezed coherent
states of light entering quantum optics \cite{15.,Celeghini:1993jh,Celeghini:1991jw} can be
identified  \cite{Celeghini:1989qc}, up
to elements of the group \ggg ~ of automorphisms of $su(1,1)$, with the
states of the quantum dho.

It has been also shown \cite{Celeghini:1992yv} that the dho states
are time dependent
thermal states, as expected due to the statistical nature of dissipation.
This is reported in Sec. 3.
The formalism
for the dho turns out to be similar to the one of real time
QFT at finite temperature,
also called thermo-field dynamics (TFD) \cite{11.,16.,17.}.
In refs. \cite{18.}  and \cite{Tsue:1994nz} such
a connection with TFD has been further analyzed and the master
equation has been discussed \cite{18.}.

In ref. \cite{Srivastava:1995yf}
the exact action for the dho in the path integral
formalism of Schwinger and Feynman and Vernon has been obtained.
The initial values of the doubled variables have been related
to the probability of quantum fluctuations in the vacuum,
a result which is interesting also in the more general case
of thermal field theories. I report such results in Sec. 4.

In Sec. 5 I  show that the proper algebraic structure of QFT
is the Hopf algebra \cite{Celeghini:1993jh}, which includes the usually
considered
Weyl-Heisenberg algebra (WH). I then show \cite{Celeghini:1998sy}
that dissipative
systems are properly described in the frame of the q-deformed
Hopf algebra \cite{25.,Celeghini:1991km,20.}.
The
q-deformation parameter turns out to be related with time
parameter in the
case of dho and  with temperature in the case of
thermal field theory. In both cases, the q-parameter acts
as a label for the ui representations.
Such a conclusion confirms
a general analysis \cite{Iorio:1994jk} which shows that the Weyl
representations
in QM and the ui representations in
QFT are indeed labeled by the deformation parameter.

Sec. 6 is devoted to the conclusions. There I mention some recent
developments which point to the r\^ole of dissipation in the
quantization procedure \cite{'tHooft:1999gk,Blasone:2001ew} and I
also shortly recall
the r\^ole of dissipation in the quantum
model of the brain \cite{Vitiello:1995wv,31.}
and on the occurrence that the cosmological
arrow of time, the thermodynamical one and the biological one
point into the same direction \cite{Alfinito:2000ck,37.}.

\bigskip

\bigskip
{\bf 2. The damped
harmonic oscillator}
\bigskip

In this section I want to perform the canonical quantization of
the damped harmonic oscillator with classical equation
\be\lab{(2.1)}
m \ddot x + \gamma \dot x + \kappa x = 0 \quad .
\ee

In the following I closely follow the approach of refs.
\cite{Celeghini:1989qc}-\cite{Celeghini:1992yv} and \cite{13.}.
The canonical
quantization scheme can only deal with an isolated system. It is then
necessary to double the phase-space
dimensions \cite{12., 13.} in order to {\it close} the system
(\ref{(2.1)}).
The closed system Lagrangian is then written as
\be\lab{(2.2)}
L = m \dot x \dot y + \half \gamma ( x \dot y - \dot x y ) - \kappa x y
\quad .
\ee

Eq. (\ref{(2.1)}) is obtained by varying eq. (\ref{(2.2)}) with
respect to $y$, whereas
variation with respect to $x$ gives
\be\lab{(2.3)}
m \ddot y - \gamma \dot y + \kappa y = 0 \quad ,
\ee
which appears to be the {\it time reversed} ($\gamma
\rightarrow - \gamma$) of eq. (\ref{(2.1)}).
$y$ may be thought of as describing an
effective degree of freedom for the heat bath to which the system
(\ref{(2.1)}) is coupled.
The canonical momenta
are then given by ${\,
p_{x} \equiv {{\partial L}\over{\partial \dot x}} = m \dot y - \half
\gamma y}$ ; ${p_{y} \equiv {{\partial L}\over{\partial \dot y}}
= m \dot x + \half \gamma x}$.
The Hamiltonian is
\be\lab{(2.4)}
H = p_{x} \dot x + p_{y} \dot y - L = {{1}\over{m}} p_{x} p_{y} +
{{1}\over{2m}}\gamma
\left ( y p_{y} - x p_{x} \right ) + \left ( \kappa - {{\gamma^{2}}\over{4 m}}
\right ) x y \quad .
\ee
For a discussion of Hamiltonian systems of this kind see also
\cite{Banerjee:2001yc}.
Canonical quantization is performed by introducing the
commutators
$[ x , p_{x} ]= i\, \hbar = [ y , p_{y} ] , ~
[ x , y ] = 0 = [ p_{x} , p_{y} ]$, and the corresponding sets of
annihilation and creation operators

\be\lab{(2.5)}
\alpha  \equiv
\left ({1\over{2 \hbar \Omega}} \right )^{1\over{2}} \left (
{{p_{x}}\over{\sqrt{m}}} - i \sqrt{m} \Omega x \right ) \quad , \quad
\alpha^{\dagger} \equiv \left ({1\over{2 \hbar \Omega}} \right )^{1\over{2}}
\left ( {{p_{x}}\over{\sqrt{m}}} + i \sqrt{m} \Omega x \right ) \quad ,
\ee

\be\lab{(2.5a)}
\beta \equiv \left ({1\over{2 \hbar \Omega}} \right )^{1\over{2}} \left (
{{p_{y}}\over{\sqrt{m}}} - i \sqrt{m} \Omega y \right ) \quad , \quad
\beta^{\dagger} \equiv \left ({1\over{2 \hbar \Omega}} \right )^{1\over{2}}
\left ( {{p_{y}}\over{\sqrt{m}}} + i \sqrt{m} \Omega y \right ) \quad ,
\ee

\be\lab{(2.5b)}
[\, \alpha , \alpha^{\dagger} \,] = 1 = [\, \beta , \beta^{\dagger} \, ] \quad
, \quad  [\, \alpha , \beta \,] = 0 = [\, \alpha , \beta^{\dagger} \, ] \quad .
\ee
\noindent I have introduced ${\Omega \equiv \left [
{1\over{m}} \left ( \kappa - {{\gamma^{2}}\over{4 m}} \right )
\right ]^{1\over{2}}}$, the common frequency of the two oscillators
eq. (\ref{(2.1)})
and eq. (\ref{(2.3)}), assuming $\Omega$ real, hence ${ \kappa >
{{\gamma^{2}}\over{4 m}}}$ (case of
no overdamping).
The Feshbach
and Tikochinsky  \cite{13.} quantum Hamiltonian is then obtained as
\be\lab{(2.6)}
H = \hbar\Omega(\alpha^{\dagger}\beta + \alpha \beta^{\dagger})
- {i\hbar\gamma\over{4m}}\left[(\alpha^{2} - \alpha^{\dagger 2})
 - (\beta^{2} - \beta^{\dagger 2})\right]  ~~.
\ee

In Sec. 4 I show that, at quantum level, the $\beta$ modes
allow quantum noise effects arising from the imaginary part
of the action \cite{Srivastava:1995yf}. In Sec. 5 the
doubling of the degrees of freedom will be shown to be
a quite natural operation implied by
the physically unavoidable requirement of the additivity of
basic observables such
as the energy, the angular momentum, etc..

By using the canonical linear transformations
$
{A \equiv {1\over{\sqrt 2}}
( \alpha + \beta )}
,$~
$
{B \equiv {1\over{\sqrt 2}}
( \alpha - \beta )},~
$
$H$ is written as
\be\lab{(2.71)}
 H =  H_{0} +  H_{I} \quad ,
\ee
\be\lab{(2.7)}
 H_{0} = \hbar \Omega ( A^{\dagger} A - B^{\dagger} B ) \quad , \quad
H_{I} = i \hbar \Gamma ( A^{\dagger} B^{\dagger} - A B ) \quad ,
\ee
where the decay constant for the classical variable $x(t)$ is
denoted by $\Gamma \equiv {{\gamma}\over{2 m}}$.

I observe that {\it the states generated by $B^{\dagger}$ represent
the sink where the energy
dissipated by the quantum damped oscillator flows: the $B$-oscillator
represents the reservoir or heat bath coupled to the
$A$-oscillator}.

The dynamical group structure associated
with the system of coupled quantum oscillators is that of $SU(1,1)$.
The
two mode realization of the algebra $su(1,1)$ is indeed generated by
$
J_{+} = A^{\dagger} B^{\dagger} , \quad J_{-} = J_{+}^{\dagger} = A B
, \quad
J_{3} = {1\over{2}} (A^{\dagger} A + B^{\dagger} B + 1) ,~
$
$
[\, J_{+} , J_{-}\, ] = - 2 J_{3} , \quad [\, J_{3}  , J_{\pm}\, ]
= \pm
J_{\pm} .$ The Casimir operator ${\cal C}$ is
${{\cal C}^{2} \equiv {1\over {4}} + J_{3}^{2} - {1\over{2}} (
J_{+} J{-} + J_{-} J_{+} )}$ ${= {1\over{4}} ( A^{\dagger} A -
B^{\dagger} B)^{2}} $.

I also observe that
$
[\, H_{0} , H_{I}\, ] = 0
$.
The time evolution of the vacuum
$
|0> \equiv | n_{A} = 0 , n_{B} = 0 > ~,~
A |0> = 0 = B |0>~
$,
is controlled by $H_{I}$
$$
|0(t)> = \exp{ \left ( - i t {H \over{\hbar}} \right )} |0> =
\exp{ \left ( - i t {H_{I} \over{\hbar}} \right )} |0>
$$
\be\lab{(2.8a)}
= {1\over{\cosh{(\Gamma t)}}} \exp{
\left ( \tanh {(\Gamma t)} A^{\dagger} B^{\dagger} \right )}|0> \quad ,
\ee

\be\lab{(2.8b)}
<0(t) | 0(t)> = 1~ \quad \forall t~,
\ee
\be\lab{(2.9)}
\lim_{t\to \infty} <0(t) | 0> \, \propto \lim_{t\to \infty}
\exp{( - t  \Gamma )} = 0 \quad .
\ee

Notice that once one sets the initial condition of positiveness
for the eigenvalues of $H_{0}$,
such a condition is preserved by the time evolution since $H_{0}$ is
the Casimir operator (it commutes with $H_{I}$). In other words, there
is no danger of dealing with energy spectrum unbounded from below.
Time evolution for creation and annihilation operators is given by
\be\lab{(2.10a)}
A \mapsto A(t) =
{\rm e}^{- i {t\over{\hbar}} H_{I}}
A ~{\rm e}^{i {t\over{\hbar}} H_{I}} =
A \cosh{(\Gamma t)} - B^{\dagger} \sinh{(\Gamma t)} ~,
\ee
\be \lab{(2.10b)}
B \mapsto B(t) =
{\rm e}^{- i {t\over{\hbar}} H_{I}}
B ~{\rm e}^{i {t\over{\hbar}} H_{I}} =
B \cosh{(\Gamma t)} - A^{\dagger} \sinh{(\Gamma t)}
\quad
\ee
and h.c., and the corresponding ones for $A(t)$, $B(t)$ and h.c..
I note that eqs. (\ref{(2.10a)}) and (\ref{(2.10b)}) are Bogolubov
transformations: they are
canonical transformations preserving the
ccr.
Eq. (\ref{(2.9)}) expresses the
instability (decay) of the vacuum under the evolution
operator $\exp{ \left ( - i t {H_{I} \over{\hbar}} \right )}$.
{\it This means that
the QM framework is not suitable for the canonical quantization
of the dho}. In other words time evolution leads out of the Hilbert space of
the states
and
in ref. \cite{Celeghini:1992yv} it has been shown that the proper way to perform
the canonical quantization of the dho is to work in the framework
of QFT. In fact
for many degrees of freedom the time evolution operator ${\cal U}(t)$
and the vacuum are formally (at finite volume) given by
\be
{\cal U}(t) =
\prod_{\kappa}{\exp\Bigl(-{\Gamma_{\kappa} t \over{ 2}}\bigl(
{\alpha}_{\kappa}^2 -
{\alpha}_{\kappa}^{\dagger 2}\bigr)
\Bigr)
\exp\Bigl({\Gamma_{\kappa} t \over{ 2}}\bigl(
{\beta}_{\kappa}^2 -
{\beta}_{\kappa}^{\dagger 2}\bigr)
\Bigr)}
\ee
\be\lab{(2.11)}
=\prod_{\kappa}{\exp{\Bigl ( \Gamma_{\kappa} t \bigl ( A_{\kappa}^{\dagger}
B_{\kappa}^{\dagger} - A_{\kappa} B_{\kappa} \bigr ) \Bigr )}},
\ee
\be\lab{(2.12)}
|0(t)> = \prod_{\kappa} {1\over{\cosh{(\Gamma_{\kappa} t)}}} \exp{
\left ( \tanh {(\Gamma_{\kappa} t)} A_{\kappa}^{\dagger}
B_{\kappa}^{\dagger} \right )} |0> \quad ,
\ee
\noindent with
$<0(t) | 0(t)> = 1~, \quad \forall t $~.
Using the continuous limit relation $
\sum_{\kappa} \mapsto {V\over{(2 \pi)^{3}}} \int \! d^{3}{\kappa}$,
in the
infinite-volume limit we have (for $\int \!
d^{3} \kappa ~
\Gamma_{\kappa}$ finite and positive)
\be\lab{(2.13)}
{<0(t) | 0> \rightarrow 0~~ {\rm as}~~ V\rightarrow \infty }
~~~\forall~  t~  ,
\ee
and in general,
$
{<0(t) | 0(t') > \rightarrow 0~~ {\rm as}~~ V\rightarrow \infty}
~~~\forall~t$~ and~ $t'~ ,~~~ t' \neq t.
$
At each time {\it t} a representation
$\{ |0(t)> \}$ of the ccr is defined and turns out to be ui
to any other
representation $\{ |0(t')>~,~~\forall t'\neq t \}$ in the infinite volume
limit. In such a way the quantum dho evolves in time through ui
representations of ccr ({\it tunneling}).
I remark that  $| 0(t)>$ is a two-mode time dependent
generalized coherent state \cite{22.,23.}.

One thus see that the Bogolubov transformations,
eqs. (\ref{(2.10a)}) and (\ref{(2.10b)})
can be implemented for every $\kappa$ as inner automorphism for the
algebra  ${su(1,1)}_{\kappa}$. At each time  {\it t}
we have a copy
$\{ A_{\kappa}(t) , A_{\kappa}^{\dagger}(t) , B_{\kappa}(t) ,
B_{\kappa}^{\dagger}(t) \, ; \, | 0(t) >\, |\, \forall {\kappa} \}$
of the
original algebra induced by the time evolution operator which
can thus be thought of as a generator
of the group of automorphisms of ${\bigoplus_{\kappa}
su(1,1)_{\kappa}}$ parameterized by time  {\it t} (we have a
realization of the operator algebra at
each time t,
which can be implemented by Gel'fand-Naimark-Segal construction in the
C*-algebra formalism \cite{10.}).
Notice that the various copies
become unitarily inequivalent in the infinite-volume limit, as
shown by eqs. (\ref{(2.13)}): the space of the states splits
into ui representations of the ccr each one labeled by time
parameter  {\it t}.
As usual, one works at finite volume and only at the end
of the computations the limit $V \to \infty$ is performed.

\bigskip

\bigskip
{\bf 3. Thermal features of quantum dissipation }
\bigskip

In refs. \cite{Celeghini:1990gr} and \cite{Celeghini:1992yv} it has been shown that the
representation $\{|0(t)>\}$ is equivalent to the TFD representation
$\{|0(\beta(t)>\}$, thus recognizing the relation between the dho
states and the finite temperature states.
In particular, one may introduce the { \it free energy}
functional for the $A$-modes
\be\lab{(2.14)}
{\cal F}_{A} \equiv <0(t)| \Bigl (  H_{A} - {1\over{\beta}} S_{A}
\Bigr ) |0(t)> \quad ,
\ee
where $H_{A}$ is the part of $H_{0}$ relative to  $A$-
modes only,
namely $H_{A} = \sum_{\kappa} \hbar \Omega_{\kappa}
A_{\kappa}^{\dagger} A_{\kappa}$, and the {\it entropy} $S_{A}$
is given by
\be\lab{(2.15)}
 S_{A} \equiv - \sum_{\kappa} \Bigl \{ A_{\kappa}^{\dagger} A_{\kappa}
\ln \sinh^{2} \bigl ( \Gamma_{\kappa} t \bigr ) - A_{\kappa}
A_{\kappa}^{\dagger} \ln \cosh^{2} \bigl ( \Gamma_{\kappa} t \bigr ) \Bigr \}
\quad .
\ee
One then considers
the stability condition
${{\partial {\cal F}_{A}}\over{\partial \vartheta_{\kappa}}} = 0 \quad
 \forall \kappa \quad ,\vartheta_{\kappa} \equiv \Gamma_{\kappa} t$~
to be satisfied in each representation,
and using the definition $E_{\kappa} \equiv \hbar \Omega_{\kappa}$, one finds
\be
%\lab{()}
{\cal N}_{A_{\kappa}}(t) = \sinh^{2} \bigl ( \Gamma_{\kappa} t \bigr ) =
{1\over{{\rm e}^{\beta (t) E_{\kappa}} - 1}} \quad , \lab{(2.16)}
\ee
namely the Bose distribution for $A_{\kappa}$ at time
{\it t}.
$\{ |0(t)> \}$ is  thus recognized to be a representation of
the ccr at finite temperature, equivalent
to the TFD representation $\{ |0(\beta)> \}$~ \cite{11.,16.,17.}.
I also notice that ${H}_{0}$ and $H_{I}$ in eq. (\ref{(2.7)})
are the free Hamiltonian and the generator of Bogolubov
transformations, respectively, also in TFD (provided one sets $\Gamma t \equiv
 \theta (\beta) $ and $\Omega$ is given a proper expression).
Use of eq. (\ref{(2.15)}) shows that
\be\lab{()}
{{\partial}\over{\partial t}} |0(t)> =  - \left ( {1\over{2}}
{{\partial {\cal S}}\over{\partial t}} \right ) |0(t)> \quad . \lab{(2.17)}
\ee
One thus see
that $i \left ( {1\over{2}} \hbar {{\partial
{\cal S}}\over{\partial t}} \right )$ is the
generator of time translations,
namely
time evolution
is controlled by the entropy variations \cite{DeFilippo:1977bk}.
It is remarkable that the same
dynamical variable
${\cal S}$
whose expectation value is formally the entropy also
controls time evolution: Damping
(or, more generally, dissipation) implies indeed the choice of a privileged
direction in time evolution ({\it
arrow of time}) with a consequent breaking of
time-reversal invariance.
One may also show that
$
d {\cal F}_{A} = d E_{A} - {1\over{\beta}} d {\cal S}_{A}=0~,
$ which
expresses the
first principle of thermodynamics for a system coupled with environment
at constant temperature and in absence of mechanical work.
One may define as usual heat as ${dQ={1\over{\beta}} dS}$
and see that the change in time $d {\cal N}_{A}$ of particles
condensed in the
vacuum turns out into heat dissipation $dQ$.

It is interesting to observe that the thermodynamic arrow of time,
whose direction is defined by the increasing entropy direction, points
in the same direction of the cosmological arrow of time, namely
the inflating time direction for the
expanding Universe. This can be shown by considering indeed
the quantization of inflationary models \cite{Alfinito:2000bv} (see also \cite{Martellini:1978sm}).
The concordance between the two
arrows of time (also with the psychological arrow of time, cf. Sec. 6)
is not at all granted and is a subject of an ongoing debate (see, e.g.,
\cite{Hawking:1996ny}).

\bigskip

\bigskip
{\bf 4. Quantum noise and the doubling of the degrees of freedom }
\bigskip

Let me now ask the following question:
Does the doubling of the degrees of freedom, namely the
introduction of an ``extra coordinate'', make any sense in the
context of conventional QM?
To answer to such a question I
consider the special case
of zero mechanical resistance. Let me begins with the
Hamiltonian for an isolated particle and the corresponding density
matrix equation
\be\lab{(3.1)}
H=-(\hbar^2/2m )(\partial/\partial Q)^2 +V(Q).
\ee
\be\lab{(3.2)}
i\hbar (\partial \rho /\partial t)=[H,\rho ],
\ee
which indeed requires two coordinates (say $Q_+$ and $Q_-$). In the
coordinate representation, we have \cite{Srivastava:1995yf}
\be\lab{(3.2a)}
i\hbar (\partial/\partial t)<Q_+|\rho (t)|Q_->=
\ee
\be\lab{(3.3)}
\{
-(\hbar^2/2m)[(\partial/\partial Q_+)^2-(\partial/\partial Q_-)^2]
+[V(Q_+)-V(Q_-)]
\}<Q_+|\rho (t)|Q_->.
\ee
In terms of the coordinates $x$ and $y$, it is $
Q_{\pm}=x\pm (1/2)y$,~
and the density matrix function
$
W(x,y,t)=<x+(1/2)y|\rho (t)|x-(1/2)y>
$.
From eq. (\ref{(3.3)}) the Hamiltonian now reads
$
{H}_{0}=(p_xp_y/m)+V(x+(1/2)y)-V(x-(1/2)y),
$ with
$
p_x=-i\hbar(\partial /\partial x),
\ \ p_y=-i\hbar(\partial /\partial y),
$ which,
of course, may be constructed from the
``Lagrangian''
\be\lab{(3.4)}
{\cal L}_{0}(\dot{x},\dot{y},x,y)=m \dot{x}\dot{y}
-V(x+(1/2)y)+V(x-(1/2)y),
\ee
One has then the justification for introducing eq. (\ref{(2.2)})
at least for the case $\gamma=0$. Notice indeed that for $
V(x\pm (1/2)y)=(1/2)k(x\pm (1/2)y)$ eq. (\ref{(3.4)}) gives eq. (\ref{(2.2)})
for the case $\gamma =0$.

Next,
my task is to explore the manner in which the Lagrangian model for
quantum dissipation of refs.\cite{Celeghini:1989qc} - \cite{Iorio:1993jn}, \cite{13.} arises
from the formulation of the quantum Brownian
motion problem as described by Schwinger \cite{8.}
and by Feynman and Vernon \cite{9.}.

Let me suppose that the particle interacts with a thermal bath
at temperature $T$. The interaction Hamiltonian between the bath and
the particle is taken as
$
H_{int}=-fQ,
$
where $Q$ is the particle coordinate and $f$ is the random force on
the particle due to the bath.
In the Feynman-Vernon formalism the effective
action for the particle has the form
\be\lab{(3.5)}
{\cal A}[x,y]=\int_{t_i}^{t_f}dt{\cal L}_o(\dot{x},\dot{y},x,y)
+{\cal I}[x,y],
\ee
where ${\cal L}_o$ is defined in eq. (\ref{(3.4)}) and
\be\lab{(3.6)}
e^{(i/\hbar){\cal I}[x,y]}=
<(e^{(-i/\hbar)\int_{t_i}^{t_f}f(t)Q_-(t)dt)})_-
(e^{(i/\hbar)\int_{t_i}^{t_f}f(t)Q_+(t)dt)})_+>.
\ee
In eq. (\ref{(3.6)}) the average is with respect to the thermal bath;
``$(.)_{+}$'' denotes time ordering and ``$(.)_{-}$''
denotes anti-time ordering.
If the interaction between the bath and the coordinate $Q$
were turned off, then the operator $f$ of the bath
would develop in time according to
$f(t)=e^{iH_R t/\hbar}fe^{-iH_R t/\hbar }$ where $H_R$ is the Hamiltonian
of the isolated bath (decoupled from the coordinate $Q$).
$f(t)$ is the force operator of the bath to be used in eq. (\ref{(3.6)}).
Assuming that the particle makes contact with the bath at the
initial time $t_i$, the reduced density matrix function is
at a final time
\be\lab{(3.7)}
W(x_f,y_f,t_f)=\int_{-\infty}^{\infty }dx_i\int_{-\infty}^{\infty }dy_i
K(x_f,y_f,t_f;x_i,y_i,t_i)W(x_i,y_i,t_i),
\ee
\be\lab{(3.8)}
K(x_f,y_f,t_f;x_i,y_i,t_i)=
\int_{x(t_i)=x_i}^{x(t_f)=x_f}{\cal D}x(t)
\int_{y(t_i)=y_i}^{y(t_f)=y_f}{\cal D}y(t)
e^{(i/\hbar){\cal A}[x,y]}.
\ee

The correlation function for the random force on the particle
is given by
$
G(t-s)=(i/\hbar )<f(t)f(s)>.
$
The retarded and advanced Greens functions are defined by
$
G_{ret}(t-s)=\theta (t-s)[G(t-s)-G(s-t)],
$
and
$
G_{adv}(t-s)=\theta (s-t)[G(s-t)-G(t-s)]~.
$
The mechanical resistance is defined $
R=lim_{\omega \rightarrow 0}{\cal R}e Z(\omega +i0^+),
$ ~with
the mechanical impedance
$Z(\zeta )$ (analytic in the upper half complex frequency
plane ${\cal I}m
~\zeta >0$) determined by the retarded Greens function
$
-i\zeta Z(\zeta )=\int_0^\infty dt G_{ret}(t)e^{i\zeta t}.
$
The time domain quantum noise in the fluctuating random force is
$
N(t-s)=(1/2)<f(t)f(s)+f(s)f(t)>.
$

The time ordered and anti-time ordered Greens functions
describe both the retarded and advanced Greens functions as well
as the quantum noise,
\be\lab{(3.9)}
G_\pm (t-s)=\pm (1/2)[G_{ret}(t-s)+G_{adv}(t-s)]
+(i/\hbar )N(t-s).
\ee

The interaction between the bath and the particle is evaluated by following
Feynman and Vernon and we find \cite{Srivastava:1995yf} for the real and the imaginary
part of the action
\be\lab{(3.10a)}
{\cal R}e{\cal A}[x,y]=\int_{t_i}^{t_f}dt{\cal L},
\ee
\be\lab{(3.10b)}
{\cal L}=m \dot{x}\dot{y}-[V(x+(1/2)y)-V(x-(1/2)y)]
+(1/2)[xF_y^{ret}+yF_x^{adv}],
\ee
\be\lab{(3.10c)}
{\cal I}m{\cal A}[x,y]=
(1/2\hbar )\int_{t_i}^{t_f}\int_{t_i}^{t_f}dtdsN(t-s)y(t)y(s),
\ee
respectively,
where the retarded force on $y$ and the advanced force on $x$ are
defined as
$
F_y^{ret}(t)=\int_{t_i}^{t_f}dsG_{ret}(t-s)y(s),~~
F_x^{adv}(t)=\int_{t_i}^{t_f}dsG_{adv}(t-s)x(s). $

Eqs. (\ref{(3.10a)}) - (\ref{(3.10c)}) are {\it rigorously exact} for
linear passive damping
due to the bath when the path integral eq. (\ref{(3.8)}) is employed for the
time development of the density matrix.

I therefore conclude that
the lagrangian eq. (\ref{(2.2)}) can be viewed as the approximation
to eq. (\ref{(3.10b)}) with $F_y^{ret}=\gamma\dot{y}$ and
$F_x^{adv}=-\gamma\dot{x}$.

I also observe that at the classical level
the ``extra'' coordinate $y$, is usually constrained to vanish.
Note that $y(t)=0$ is a true solution to
eqs. (\ref{(2.3)}) so that the constraint is {\it
not} in violation of the equations
of motion.
From eqs. (\ref{(3.10a)}) - (\ref{(3.10c)}) one sees that
{\it at quantum level nonzero $y$ allows quantum
noise effects arising from the imaginary part of the action}.
On the contrary, in the classical ``$\hbar \rightarrow 0$'' limit
nonzero $y$ yields an
``unlikely process'' in view of the large imaginary part of the action
implicit in
eq. (\ref{(3.10c)}). Thus, the meaning of
the constraint $y=0$ at the classical level is the one of avoiding such
``unlikely process''.

$$   $$

\bigskip

\bigskip
\noindent {\bf 5. Hopf algebra, q-deformation and quantum dissipation}
\bigskip

Quantum deformations \cite{25.,20.}
of Lie algebras are well studied mathematical
structures and therefore their properties need not to be presented
again in this paper. I only recall that they are
deformations of the enveloping algebras of Lie algebras and have
Hopf algebra structure \cite{Celeghini:1991km}.
In this Section I will show \cite{Celeghini:1998sy} that dissipative
systems (as well as the finite temperature non-equilibrium
systems) are properly described in the frame of the q-deformed
Hopf algebra. Moreover,
I will argue that the
the proper algebraic structure of QFT
is the deformed Hopf algebra.
The
q-deformation parameter turns out to be related with the time
parameter in the
case of dho (and  with temperature in the case of
thermal field theory). In both cases, the q-parameter acts
as a label for the ui representations.

I observe that one central ingredient of Hopf
algebras is the operator doubling
implied by the coalgebra. The coproduct operation is indeed a map ${\Delta}:
{\cal A}\to {\cal A}\otimes {\cal A}$ which duplicates the algebra.
{\it Lie-Hopf
algebras are commonly used in the familiar addition of energy, momentum and
angular momentum}, e.g., for the $''$addition$\, ''$ of the angular momentum
$J^{\alpha}$, ${\alpha} = 1,2,3$, of two particles one has:
$\Delta J^{\alpha} = J^{\alpha}  \otimes {\bf 1} + {\bf 1} \otimes
J^{\alpha} \equiv J^{\alpha}_1 + J^{\alpha}_2,~ J^{\alpha} \in su(2)$.
Thus, the physical meaning of the coproduct is
that it provides the prescription
for operating on two modes.

In the following, for simplicity, let me focus on the case of bosons.
The conclusions can also be extended to fermions \cite{Celeghini:1998sy}.

The bosonic Hopf algebra, also called $h(1)$, is  generated by the
set of operators $\{ a,
a^{\dagger},H,N \}$ with commutation relations:
\be
[\, a\, ,\, a^{\dagger} \, ] = \ 2H \, , \quad\;
[\, \ N \, ,\, a \, ] = - a \, , \quad\; [\, \ N \, ,\, a^{\dagger} \, ] =
a^{\dagger} \, , \quad\; [\, \ H \, ,\, \bullet \, ] = 0 \, .
\lab{p22}
\ee
Here $a$ and $a^{\dagger}$ denote generic
annihilation and creation operators. For notational simplicity
I omit the momentum suffix $\kappa$ which will be restored later on.
Later we will see how the present
discussion relates to the dho operators introduced in the previous
Sections.
$H$ is a central operator, constant in each representation. The Casimir
operator is given by ${\cal C} = 2NH -a^{\dagger}a$.
~$h(1)$ is equipped with the coproduct
operation, defined by
\be
\Delta a = a \otimes {\bf 1} + {\bf 1} \otimes a \equiv a_1 + a_2 ~,~~~
~\Delta a^{\dagger} = a^{\dagger} \otimes {\bf 1} + {\bf 1} \otimes
a^{\dagger} \equiv a_1^{\dagger} + a_2^{\dagger} ~,
\lab{p23}\ee
\be
\Delta H = H \otimes {\bf 1} + {\bf 1} \otimes H  \equiv H_{1} +
 H_{2}~, ~~~\Delta N = N \otimes {\bf 1} + {\bf 1} \otimes N \equiv  N_{1} +
 N_{2} ~.  \lab{p24}
 \ee

I remark that usually one introduces the
operator algebra necessary to set up QFT by limiting himself
to the introduction of the boson
Weyl-Heisenberg (WH) algebra (\ref{p22}).
The assumption of the additivity of some observables such as the energy,
the momentum and the angular momentum is so obvious that one does not
even bother to spell it out. It is implicitly given as granted. However,
if one is asked to express it explicitly and formally, then it becomes
natural to introduce the coproduct map, as shown above, and thus to
realize that the boson WH algebra (\ref{p22}) is only a part
of the full algebraic structure. One needs the Hopf structure. The
full algebraic structure which is needed, however, has to take into account
one of the very special features of QFT,
the one which characterizes it and makes it different from QM,
namely the existence of infinitely many representations of the ccr
(in QM all the representations of the ccr are unitary equivalent due
to the von Neumann theorem).
Then one is led to consider the quantum deformation of the Hopf algebra,
as it appears from the following.

The $q$-deformation of $h(1)$ is the Hopf algebra $h_{q}(1)$:
\be
[\, a_{q}\, ,\, a_{q}^\dagger \, ] = \ [2H]_{q} \, , \quad\;
[\, \ N \, \, , \, a_{q}\, ] = - a_{q} \, , \quad\;  [ \, \ N \, , \,
a_{q}^\dagger \,] = a_{q}^\dagger , \quad\; [\, \ H \, \, , \,  \bullet \,
] = 0  \, , \lab{p26}
\ee
where $N_{q} \equiv N$ and $H_{q} \equiv H$.  The Casimir operator ${\cal
C}_{q}$ is given by ${\cal C}_{q} = N[2H]_{q} -a_{q}^{\dagger}a_{q}$, where
$\displaystyle{[x]_{q} = {{q^{x} - q^{-x}} \over {q - q^{-1}}}}$.
The coproduct is defined by
\be
\Delta a_{q} = a_{q} \otimes {q^{H}} + { q^{-H}} \otimes a_{q}
\, , \quad\quad \Delta a_{q}^{\dagger} = a_{q}^{\dagger} \otimes {q^H} +
{q^{-H}} \otimes a_{q}^{\dagger} ~, ~\lab{p28}
\ee
\be
\Delta H = H \otimes {\bf 1} + {\bf 1} \otimes H ~,~~~~ \Delta N
= N \otimes {\bf 1} + {\bf 1} \otimes N ~, \lab{p29}
\ee
whose algebra of course is isomorphic with (\ref{p26}): ~$ [ \Delta a_{q} ,
\Delta a_{q}^{\dagger} ] = [2 {\Delta} H]_{q}$ , etc. .
Note that $h_{q}(1)$ is a structure different from the commonly
considered
$q$-deformation of the harmonic oscillator~\cite{20.} that does not have a
coproduct (and thus cannot allow for the duplication of space).

Let me denote by ${\cal F}_{1}$  the single mode Fock space,
i.e. the fundamental
representation $H = 1/2$, ${\cal C} = 0$. In such a representation $h(1)$ and
$h_{q}(1)$ coincide as it happens for $su(2)$ and $su_{q}(2)$ for the
spin-$\frac{1}{2}$ representation. The differences appear in the coproduct
and in the higher spin representations.

As customary, one requires that $a$ and $a^{\dag}$, and $a_{q}$ and
${a_{q}}^{\dag}$,  are adjoint operators. This implies that $q$ can only be
real or of modulus one.
In the two mode Fock space ${\cal F}_{2} = {\cal F}_{1} \otimes {\cal F}_{1}$,
for $|q|=1$, the hermitian conjugation of the coproduct must be supplemented
by the inversion of the two spaces for consistency with the coproduct
isomorphism.

Summarizing,  on ${\cal F}_{2}  =
{\cal F}_{1} \otimes {\cal F}_{1}$ it can be written:
\be
\Delta a =  a_1 + a_2 ~,~~~ ~\Delta a^{\dagger} = a_1^{\dagger} +
a_2^{\dagger} ~, \lab{p212}
\ee
\be
\Delta a_{q} =  a_1 q^{1/2} + q^{-1/2} a_2 ~,~~~
~\Delta a_{q}^{\dagger} = a_1^{\dagger} q^{1/2}  +q^{-1/2}  a_2^{\dagger} ~,
\lab{p213}
\ee
\be
\Delta H = 1 , ~~~\Delta N =  N_{1} +
 N_{2} ~.  \lab{pdelta} \lab{p214}
\ee

Note that $[a_i , a_j ] = [a_i , a_{j}^{\dagger} ] = 0 ,
~ i \neq j $.

It is now possible
to show that the full set of infinitely many unitarily
inequivalent representations of the ccr in  QFT are classified by use
of the deformed Hopf algebra. To do that it is sufficient
to show that the Bogolubov
transformations are directly obtained by use of the deformed copodruct
operation. As well known, indeed, the Bogolubov transformations relate
different (i.e. unitary inequivalent) representations.
I consider therefore
the following operators (cf. (\ref{p28})
with $H=1/2$):
\be
{\alpha}_{q(\theta)} \equiv { { {\Delta} a_{q}} \over {\sqrt{[2]_{q}} }} =
{1\over\sqrt{[2]_{q}}} (e^{\theta} a_1 +
e^{-\theta} a_2 ) ~, \lab{p310}
\ee
\be
{\beta}_{q(\theta)} \equiv { 1 \over {\sqrt{[2]_{q}}} } {\delta \over
{\delta \theta}} {\Delta} a_{q} = {2q \over \sqrt{[2]_{q}}}{{\delta}\over
{\delta q}} \Delta a_{q} = {1\over\sqrt{[2]_{q}}} (e^{\theta}
a_1 -e^{-\theta} a_2 ) \; ,
\lab{p311}
\ee
and h.c., with $q(\theta) \equiv e^{2\theta}$  .
A set of commuting
operators with canonical commutation relations is given
by
\be
{\alpha}(\theta) \equiv {{\sqrt{[2]_{q}}}  \over 2{\sqrt2}}
[ {\alpha}_{q(\theta)} +
{\alpha}_{q(- \theta)} - {\beta}_{q(\theta)}^{\dagger}  +  {\beta}_{q(- \theta)}^{\dagger}] ~,
\lab{p314}
\ee
\be
{\beta}(\theta) \equiv {{\sqrt{[2]_{q}}}  \over 2{\sqrt 2}}[{\beta}_{q(\theta)} +
{\beta}_{q(- \theta)} - {\alpha}_{q(\theta)}^{\dagger}  +  {\alpha}_{q(- \theta)}^{\dagger} ] ~.
\lab{p315}
\ee
and h.c.
One then introduces
\bea
A(\theta) \equiv \frac{1}{\sqrt{2}} \left ( {\alpha}(\theta ) +
{\beta}(\theta )\right )
&=& A ~{\rm cosh} ~\theta - {B}^{\dagger}
~{\rm sinh} ~\theta ~~, \lab{p321a} \\
B(\theta) \equiv \frac{1}{\sqrt{2}} \left ( {\alpha}(\theta ) -
{\beta}(\theta )
\right ) &=& B ~{\rm cosh} ~\theta -
A^{\dagger} ~{\rm sinh} ~\theta ~, \lab{p321}
\eea
with
\be
[ A(\theta) , A^{\dagger}(\theta) ] = 1 ~, ~~[ B(\theta) ,
B^{\dagger}(\theta) ] = 1 ~.~~
\lab{p322}
\ee
All other commutators are equal to zero and $A(\theta)$ and
$B(\theta)$ commute among themselves.

Eqs. (\ref{p321a}) and (\ref{p321})  are nothing
but the Bogolubov transformations for the $(A,B)$  pair, to be compared
with the corresponding transformations (\ref{(2.10a)}) and (\ref{(2.10b)})
in the case of the dho.
In other words, eqs. (\ref{p321a}),
(\ref{p321}) show that the Bogolubov-transformed operators $A(\theta)$ and
$B(\theta)$ are linear combinations of the coproduct operators
defined in terms of the deformation parameter $q(\theta )$ and of their
${\theta}$-derivatives.

From this point on one can re-obtain the results discussed in
the previous Sections for the dho, provided one sets
$\theta \equiv \Gamma t$.
Notice that
\be
- i{\delta \over {\delta \theta}} A(\theta) =
[{\cal G}, A(\theta)] ~,~~~
- i{\delta \over {\delta \theta}} B(\theta) =
[{\cal G}, B(\theta)] ~,
\lab{p326}\ee
and h.c., where ${\cal G} \equiv -i(A^{\dagger}B^{\dagger} - AB)$
denotes the generator of (\ref{p321a}) and
(\ref{p321}).
For a fixed value $\bar{\theta}$, we have
\be
\exp(i{\bar{\theta}} p_{\theta}) ~A(\theta) =
\exp(i{\bar{\theta}}{\cal G}) ~A(\theta)~
\exp(-i{\bar{\theta}}{\cal G}) = A( \theta + {\bar {\theta}} )
~,
\lab{p327}\ee
and similar equations for $B(\theta)$.

In eq.(\ref{p327}) the definition $\displaystyle{p_{\theta}
\equiv -i{\delta \over {\delta \theta}}}$ has been used. It
can be regarded as the
momentum operator $''$conjugate$\, ''$ to the $''$degree of
freedom$\, ''$ $\theta$, which thus
acquires formal definiteness in the sense of the canonical
formalism.
In the infinite volume limit $<0(\theta)|0(\theta ')> = 0$.
In other words, the deformation parameter
$\displaystyle{\theta = {1 \over {2}} ~{\rm ln}~q}$
acts as a label for the inequivalent representations,
consistently with the results of Refs.~\cite{Iorio:1993jn,Celeghini:1993jh}.
It is remarkable that the "conjugate momentum" $p_{\theta}$
generates transitions among inequivalent (in the infinite volume limit)
representations: $\exp (i{\bar \theta}p_{\theta})~ ~|0(\theta)> = |0(\theta +
{\bar \theta})>$.

In conclusion, one obtains, by use of the deformed Hopf algebra,
the typical structure one deals with in QFT.
In this connection, I observe that variation in time of the deformation
parameter is related with the so-called heat-term in dissipative systems. In
such a case, in fact, the Heisenberg equation for $A(t,{\theta}
(t))$ is
$$
-i{\dot A}(t,{\theta}(t)) = -i{\delta \over {\delta t}} A(t,{\theta}(t))
-i{{\delta \theta} \over {\delta t}}~ {\delta  \over {\delta \theta}}A(t,
{\theta}(t))=
$$
%\nonumber \\
\be
\left [ H , A(t,{\theta}(t)) \right ] +
{{\delta \theta} \over {\delta t}}~ [{\cal G}, A(t,{\theta}(t)) ] =
\left [ H + Q , ~A(t,{\theta}(t)) \right ] ~, \lab{42}
\ee
where $\displaystyle{Q \equiv {{\delta \theta} \over {\delta t}} {\cal G}}$
denotes the heat-term, and $H$ is the Hamiltonian
(responsible for the time variation in the explicit time dependence of
$A(t,{\theta}(t))$).  $H + Q$ is therefore to be identified with the
free energy~\cite{Celeghini:1992yv,DeFilippo:1977bk}.
In this way the results of
Sec. 3 are also recovered.
Thus, the conclusion is that variations in time of the deformation
parameter actually involve dissipation.

When the proper field description is taken into account, $A$ and
$B$ carry dependence on the momentum $\kappa$ and, as customary in
QFT, one deals with the algebras $\displaystyle{
\bigoplus_{\kappa} h_{\kappa}(1)}$ (cf. Sec. 2).

\bigskip

\bigskip
{\bf 6. Concluding remarks}

\bigskip
The dho total Hamiltonian is invariant under the transformations
generated by ${J_{2} = \bigoplus_{\kappa} J_{2}^{(\kappa )}}$. The
vacuum however is not invariant under $J_{2}$ (see eq. (\ref{(2.13)}))
in the infinite
volume limit. Moreover, at each time t, the representation $\{|0(t)>\}$~
may be
characterized by the expectation value in the state $|0(t)>$  of,
{\it e.g.},
${J_{3}^{(\kappa )} - {1\over{2}}}$: thus the total number of
particles~ $n_{A} +n_{B} = 2n$~ can be taken as an order parameter. Therefore,
at each time t the symmetry under $J_{2}$ transformations is spontaneously
broken. On the other hand, ${H}_{I}$ is proportional to $J_{2}$. Thus, in
addition to the breakdown of time-reversal (discrete) symmetry, already mentioned
in Secion 2, we also have spontaneous breakdown of time
translation (continuous) symmetry.
In other words, dissipation (i.e. energy
non-conservation), has been described as an effect of the
breakdown of time translation and time-reversal symmetry. It is an interesting
question asking which is the zero-frequency mode, playing the r\^ole of
the Goldstone mode,
related with the breakdown of continuous time translation symmetry: I observe
that since~ $n_{A} - n_{B}$ is constant in time, the condensation (annihilation
and/or creation) of AB-pairs does not contribute to the vacuum energy so that
AB-pair may play the r\^ole of a zero-frequency mode.

In the discussion presented above a crucial r\^ole is
played by the existence of infinitely many ui
representations of the
ccr in QFT. In refs. \cite{Celeghini:1993jh,Iorio:1994jk} the q-WH algebra has been discussed in
relation with the von Neumann theorem in QM and it has been
shown on a general ground that
the q-deformation parameter acts as a label for the
Weyl systems in QM and for the ui
representations in QFT; the
mapping between {\it different}
 ($i.e.$ labeled by different values of $q$)
representations (or Weyl systems) being
performed by the Bogolubov transformations.
Damped
harmonic oscillator and
finite temperature systems are explicit examples clarifying
the physical meaning of such a labeling.
Further examples are provided by unstable particles in QFT \cite{DeFilippo:1977bk},
by
quantization of the matter field in curved space-time \cite{Martellini:1978sm}, by
theories with spontaneous breakdown of symmetry where  different
values of the order parameter are associated to
different ui representations (different {\it phases}).
In the case of damping,
as well as in the
case of time-dependent temperature,
the system time-evolution is represented as
{\it tunneling} through {\it
ui} representations: the non-unitary
character of time-evolution ({\it arrow of time}) is thus expressed
by the non-unitary equivalence of the representations in the infinite
volume limit. It is remarkable that at the algebraic level this is
made possible through the
q-deformation mechanism which
organizes the representations in an {\it ordered set} by means of
the labeling.

In conclusion, from the point of view of boson condensation,
time evolution in the
presence of damping may be thus thought of as a sort of continuous transition
among different phases, each phase corresponding, at time  {\it t},
to the
coherent state representation $\{|0(t)>\}$.
The damped oscillator thus provides an archetype of
system  undergoing continuous phase transition.

As already mentioned in the introduction,
dissipation in classical deterministic systems (such as the couple
of classical oscillators described by eqs. (\ref{(2.1)}) and
(\ref{(2.3)}) in Sec. 2)
has been shown \cite{Blasone:2001ew} to lead under suitable conditions to
a quantum behavior, as originally proposed by `t Hooft \cite{'tHooft:1999gk}.
In particular, dissipation
manifests as a geometric Berry-Anandan-like phase \cite{Blasone:1998xt} and it
appears to be responsible for the zero point energy contribution in the
oscillator energy spectrum \cite{Blasone:2001ew}.

The features of the dissipative quantum dynamics discussed in
this paper have been also used \cite{Vitiello:1995wv,Alfinito:2000ck,37.}
to implement an infinite memory capacity in the
quantum model of brain \cite{31.}. The key point is in the fact that
dissipative dynamics implies infinitely many degenerate vacua
(i.e. the zero eigenvalue eigenstates of $H_{0}$, eqs.
(\ref{(2.71)}) and (\ref{(2.12)})), each of them describing
a possible memory state according
to the brain model of ref. \cite{31.}. Moreover, in view of the thermal
character of such vacua illustrated in Sec. 3, the irreversible
time evolution of the brain memory states, which is perceived as
the arrow of time at a psychological experience level, appears to proceed
in the same direction of the thermodynamical and of the cosmological arrow
of time mentioned in Sec. 3. It is interesting to note that in a somewhat
unexpected way it emerges a possible answer to the questions raised by the
ongoing debate \cite{Hawking:1996ny} on the coincidence (or not) of the directions
of the three arrows of times just mentioned. Finally, the dissipative
quantum model of brain has revealed to be also interesting in the study
of features related with consciousness mechanisms \cite{37.}.

I am glad to acknowledge the Organizers of the
XXIV International Workshop on Fundamental Problems of High
Energy Physics and Field Theory, Protvino, June 2001,
and in particular Professor A.A.Logunov and Professor V.A.Petrov
for the kind and warm hospitality.
I also acknowledge for partial financial support the INFN,
INFM, MURST and the ESF Network COSLAB.

\newpage

%%%%%%%%%%%%%%%%%%%%%%%%%%%%%%%%%%%%%%

\end{document}